**ARTICLE**

# Mean Field-based Dynamic Backoff Optimization for MIMO-enabled Grant-Free NOMA in Massive IoT Networks

Haibo Wang [1], Hongwei Gao [1],[*], Pai Jiang [1], Matthieu De Mari [2], Panzer Gu[3] and Yinsheng Liu[1]

[1] School of Electronic and Information Engineering, Beijing Jiaotong University, Beijing, 100044, China

[2] Department of Information Systems Technology and Design Pillar, Singapore University of Technology and Design, Singapore, 487372, Singapore

[3] Nokia Group, Alcatel Lucent Shanghai Bell, Shanghai, 200120, China

[*]Corresponding Author: Hongwei Gao. Email: 22120049@bjtu.edu.cn

**Abstract**

In the 6G Internet of Things (IoT) paradigm, unprecedented challenges will be raised to provide massive connectivity, ultra-low latency, and energy efficiency for ultra-dense IoT devices. To address these challenges, we explore the non-orthogonal multiple access (NOMA) based grant-free random access (GFRA) schemes in the cellular uplink to support massive IoT devices with high spectrum efficiency and low access latency. In particular, we focus on optimizing the backoff strategy of each device when transmitting time-sensitive data samples to a multiple-input multiple-output (MIMO)-enabled base station subject to energy constraints. To cope with the dynamic varied channel and the severe uplink interference due to the uncoordinated grant-free access, we formulate the optimization problem as a multi-user non-cooperative dynamic stochastic game (MUN-DSG). To avoid dimensional disaster as the device number grows large, the optimization problem is transformed into a mean field game (MFG), and its Nash equilibrium can be achieved by solving the corresponding Hamilton-Jacobi-Bellman (HJB) and Fokker-Planck-Kolmogorov (FPK) equations. Thus, a Mean Field-based Dynamic Backoff (MFDB) scheme is proposed as the optimal GFRA solution for each device. Extensive simulation has been fulfilled to compare the proposed MFDB with contemporary random access approaches like access class barring (ACB), slotted-Additive Links On-line Hawaii Area (ALOHA), and minimum backoff (MB) under both static and dynamic channels, and the results proved that MFDB can achieve the least access delay and cumulated cost during multiple transmission frames.

**Keywords**: 6G; Internet of Things; grant-free random access; NOMA; dynamic backoff; mean field game

## 1. Introduction

In the 6G IoT paradigm, grant-free (GF) with non-orthogonal multiple access (NOMA) techniques is considered a key enabler for massive ultra-reliable and low-latency communication (mURLLC) services to facilitate smart transportation, smart factory, smart grid, and other mission-critical applications[1]-[3]. GF random access allows wireless terminals to transmit their preamble and data to the base station (BS) in one shot and avoid the four-handshake process in grant-based random access [4]. The combination of GF and NOMA simultaneously solves the problem of access delay, signaling overhead, as well as the scarcity of orthogonal channel resources in conventional massive access schemes [5]-[7]. Existing NOMA schemes for GF access include power-domain NOMA (PD-NOMA), code-domain NOMA, or interleave-based NOMA [8]. While PD-NOMA has been studied extensively in [5]-[7], it may introduce a long decoding delay for massive GF access devices due to the successive interference cancellation (SIC) receiver employed to distinguish different PD-NOMA signals sequentially. On the contrary, in code-domain NOMA, such as Sparse Code Multiple Access (SCMA), it allows multiple users to occupy the same resource block at the same time, achieving efficient use of spectrum resources, and SCMA uses the message passing algorithm (MPA) for detection [9]. MPA has low complexity and good performance. When multiple users access at the same time, it can effectively detect and decode users, which is crucial to support large-scale IoT device access.

Meanwhile, massive multiple-input multiple-output (MIMO) antennas are expected to be equipped on all 6G BSs. By using receiver beamforming (e.g., Zero-Forcing (ZF) [10]) at the BS, GF-NOMA transmitters can be differentiated based on their spatial characteristics, which means the access devices could be divided into multiple spatial beams (clusters)    and each preamble may be reused among multiple spatial clusters to accommodate even more access devices simultaneously [11]-[13].

In this work, we investigate the optimal backoff strategy for IoT devices in MIMO-based GF-NOMA systems within the mURLLC paradigm, applicable to scenarios such as intelligent transportation, autonomous driving, and smart factories. The proposed strategy not only effectively meets the low latency requirements of URLLC but also reduces the probability of interference between users. Additionally, it can improve the system resource allocation efficiency, thereby enhancing the overall spectrum resource utilization. With GF-NOMA, each IoT device needs to select its access parameters in a distributed manner, which will cause severe system interference and network access congestion when the number of active devices is large. Conventional ALOHA-like multiple access schemes have the devices to select a backoff time based on a random factor [14][15], which might be efficient for semi-static IoT services but far from optimal under the highly dynamic environment and the stringent delay constraint of mURLLC. Theoretically, when a large number of devices compete for limited communication resources with distributed decision-making subject to highly dynamic system states, this problem can be formulated as a DSG, and the optimal solution can be derived by solving multiple correlated stochastic differential equations (SDEs). When the amount of devices is large, it becomes prohibitively difficult to solve these SDEs simultaneously. In this work, we propose to employ mean field game (MFG) theory to solve the dynamic stochastic game (DSG) of massive IoT devices in their GF SCMA processes to minimize their average backoff delay under a limited energy budget. To the best of our knowledge, this is the first work that adopts MFG theory to dynamically optimize the backoff strategy for multi-beam MIMO based on GF-NOMA. The contributions of this study can be summarized as follows:

- A two-step GF random access scheme is proposed for MIMO BF-based cells, in which

- SCMA is adopted for multiple IoT devices within the same antenna beam, and ZF is employed to eliminate inter-beam interference in the uplink.
- We formulate a backoff delay minimization problem in GF-NOMA for mURLLC services as a multiuser non-cooperative DSG, subject to the dynamic channels, energy states, and interference among NOMA devices. In this DSG, the objective of each device is to seek the optimal dynamic backoff strategy within energy constraints to minimize the long-term backoff delay costs.
- We adopt the MFG to simplify the complex interplay between device backoff strategies. In order to obtain the optimal backoff scheme, we derive the Hamilton-Jacobi-Bellman (HJB) and Fokker-Planck Kolmogorov (FPK) equations, which are relevant to achieve the mean-field equilibrium (MFE). By solving these two coupled equation pairs iteratively with the finite difference method (FDM), we obtain the optimal backoff strategy and the evolution of the system states.
- We numerically evaluate the performance of the proposed Mean Field-based Dynamic Backoff (MFDB) scheme in comparison with conventional GF schemes based on access class barring (ACB) and slotted-Additive Links On-line Hawaii Area (ALOHA). Numerical results show that the proposed scheme can minimize the backoff delay cost and maintain a nearly constant backoff delay when the number of devices increases rapidly.

The rest of this paper is organized as follows. The related work and contributions are introduced in Section 2. The system model is presented in Section 3, and the problem formulation is described in Section 4. The MFG approach and the corresponding Dynamic Backoff Algorithm are proposed in Section 5. Section 6 numerically evaluates the performance of our proposal and other contemporary random access schemes. Finally, Section 7 concludes the paper.

## 2. Related Work

Combining GF-NOMA and beam-space MIMO can increase system capacity, improve spectral efficiency and reduce access delay, making it a promising solution for wireless communication systems. However, adopting NOMA can lead to severe co-channel interference, especially in ultra-dense IoT scenarios, where interference analysis and resource allocation become challenges. To address the above issues, the authors in [15] propose a Random Access NOMA (RA-NOMA) transmission protocol for IoT networks that employs a timer and power backoff strategy. However, this method significantly increases energy consumption. This poses a substantial negative impact on devices that require long-term operation and rely on battery power, thereby limiting the effectiveness and feasibility of this method in practical applications. The authors in [16] propose a detailed offloading protocol for the GF-SCMA enhanced MEC scheme. However, relying solely on SCMA codebooks to differentiate users in the event of resource access conflicts is insufficient, as it results in significant resource consumption for codebooks, especially with a large number of devices. In [17], the authors propose an optimization method to maximize the service quality of SCMA grant-free access with multipacket reception (MPR). In the event of a collision, the user skips the current frame with the probability of collision, and the colliding and queuing users continue to wait for the next transmission in a random time slot in another frame according to the random escape strategy. However, the random waiting time for each user after a collision is not the optimal choice for the system, potentially causing the user equipment to wait during unnecessary periods and increasing overall delay. The above MIMO-NOMA studies only consider a limited number of devices within the cell, primarily because an increase in the number of devices will lead to increased interference and the complexity of resource allocation. Besides, no works have optimized the backoff delay of a Massive SCMA-based GF-NOMA system, considering the dynamic change for system states under the limited device energy budget. To the best of our knowledge, this is the first work to propose a dynamic backoff scheme for SCMA-based GF-NOMA with practical MIMO settings.

For interference management and resource allocation in ultra-dense IoT systems, game theory can be employed to analyze the cooperation and competition among rational devices while developing strategies to maximize their payoff [18]. In the existing resource allocation schemes based on the game theory, the author of [19] proposed a power allocation framework based on cognitive radio NOMA which optimized the utility function of each device and proved the existence of Nash equilibrium. The author of [20] has proposed a Nash Bargaining Solution-based (NBS) game to achieve the optimal power allocation scheme based on channel conditions in a MIMO-NOMA system while ensuring both allocation fairness and maximum transmission rate. According to these papers, when multiple devices compete for limited communication resources in a distributed game, the dynamic optimization problem can be transformed into a DSG. However, as described by the author in [21], the device's DSG process in the ultra-dense IoT scenario will generate many SDEs, resulting in the dimensional explosion problem. To overcome the issues mentioned above, the authors of [22] proposed the MFG to transform the one-to-one interaction between devices into a more tractable interaction between the device and the mean field.

MFG is created to describe the collective behavior of a large number of interacting individuals in a system [22]. It handles the interactions in complex systems by simplifying and approximating them, and simplifies the influence of individuals on other individuals into an average effect, which helps to understand and analyze the macroscopic behavior of the system. Therefore, MFG has been widely used in optimizing the performance of large scale communication systems, which involves energy efficiency [23], transmission rate [24], and transmission power [25]. The application of MFG to the NOMA system can transform massive devices into a continuum and simulate their state distributions, thereby simplifying the complex interference into the mean field interference, which is easier to analyze. In related studies, the author of [26] proposed a NOMA-based resource allocation scheme for ultra-dense mobile edge computing (MEC) systems. To address this problem, the authors divided it into two subproblems, device clustering and power allocation. They clustered the devices based on the channel gain and proposed a resource allocation algorithm using the mean-field framework. The authors of [27] addressed the power control problem in Massive Machine Type Communication (mMTC) systems. When performing successive interference cancellation (SIC) at the receiving end, the interference is estimated by converting the location-based interference into a more manageable mean-field interference. However, SIC requires strict power ordering, and the complexity of interference estimation is greatly increased when multiple system states are considered simultaneously. Different from the previous mean-field-based power allocation schemes, in this paper, we investigate the massive GF-NOMA problem in a dynamic radio environment for the 6G IoT scenario. Our approach focuses on dynamic changes in device energy and channel state with a limited energy budget based on MFG and SCMA to minimize the backoff delay.

## 3. System Model

As shown in Fig. 1, we consider a 6G single-cell system in which a BS equips with $L$ antennas in the cell center, and $N$ ($n \in N = \{1,...,N\}$) single antenna IoT devices locate in this circular cell following a two-dimensional spatial Poisson distribution with density $\rho$. Through the fixed grid of beams (GoB), the whole cell coverage area is divided into M beams [28]. Devices with the same beam are selected to form a cluster. Considering that each radio frequency (RF) chain supports at most one device in the same time-frequency resources [29], we assume that the number of RF chains adopted at the BS is equal to the number of beams. Each RF chain provides services for devices within the corresponding beam respectively. Devices within the same cluster employ SCMA and the grant-free random access protocol for data uploading. Based on the NB-IoT standard [30], all devices in the cell share the same subcarrier and adopt time division duplexing (TDD) mode. Time $t \in \mathcal{T} = [0,T]$ is divided into frames with equal duration $\Delta t$ and the frame index is denoted by $i \in \mathcal{I} = \{1,...,I_{index}\}$ which satisfies $T = I_{index}\Delta t$. Each frame is further divided into K ($k \in \mathcal{K} = \{1,...,K\}$) time-slots (TSs) with duration $\Delta \tau$ per TS and satisfies $\Delta t = K\Delta \tau$. Assuming that the device needs to upload the status update packet periodically at each frame, whose transmission requires exactly one TS. The channel realization is described as a block-fading channel model, which remains unchanged within a frame but may vary between frames. During the packet upload process, we define the backoff delay as the time interval between the start of each frame and the data transmission TS, which can be expressed as $D_n(i) \in \{\Delta \tau, 2\Delta \tau,..., K\Delta \tau\}$. In grant-free random access (GFRA), each device needs to independently decide its backoff delay $D_n(i)$ at the beginning of each frame similar to the slotted-ALOHA protocol [31] and the transmission power $p_n(i, D_n(i))$ is adjusted indirectly based on its backoff delay and quality-of-service (QoS) constraint.

The GFRA procedure is illustrated in Fig.2, which can be divided into two main stages: broadcasting and data transmission.

*Stage I – Broadcasting:* Broadcasting: Before the beginning of each uplink transmission session within time [0, T], the BS will broadcast the pre-derived optimal MFDB policy set [32] and the statistic channel variation models all the IoT devices in this cell, as well as the available frequency resources, the trained path loss model, preamble configuration information, and reference signals. The preamble configuration information, in particular, details the format that devices must follow to generate SCMA preambles. Upon receiving these broadcast messages, each device performs channel estimation, selects an access beam based on the strength of the reference signals, and a SCMA preamble. Besides, it will predict the channel variations in the next *T* time duration based on the statistic channel variation models from the BS.

*Stage II - Data transmission:* The device can derive the channel state of each frame according to the initial channel state and the predicted channel evolution model. Before data transmission, each device selects its optimal backoff delay using our proposed MFDB scheme (the optimal policy set has been derived by the BS), according to its predicted channel states and remaining energy level. Following this backoff period, the device generates the preamble based on the configuration information received during the broadcast stage and appends it to the header of the upload packet. The detailed workings of the MFDB scheme are elaborated in Sec. A- Sec. D.

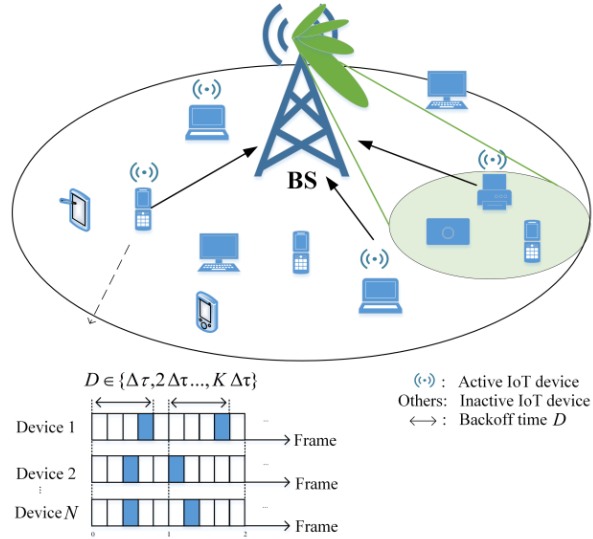

**Figure 1:** system model

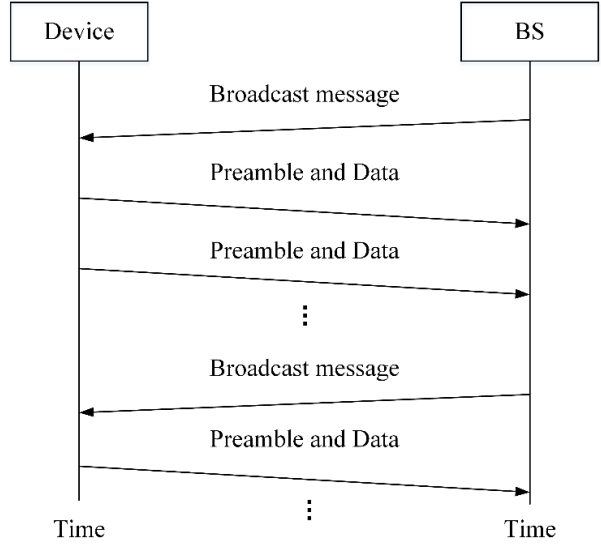

**Figure 2:** Illustration of grant free random access procedure

### 3.1. MIMO Channel Evolution

In this work, the uplink channel gain of each IoT devices is modeled with two components, namely the path-loss and the fading component. Assuming that the devices move slowly relative to the investigated transmission period, the path-loss $l_n$ will keep constant during $[0,T]$ (thus not relevant to time index $i$) and can be expressed as:

$$l_n = \min(1, \frac{1}{r_n^a}) \qquad (1)$$

where $a$ is the path loss coefficient, and $r_n$ is the distance between the device $n$ and the base station. The small fading component of device $n$ in the beam cluster $m$ is denoted as $h_{nm}(i) \in C^{L \times 1}$, modeled as an Itô process [21][33], i.e.,

$$h_{nm}(i+1) = h_{nm}(i) + \alpha_{nm}(i, h_{nm}(i))\Delta t + \sigma_{nm}(i)\Delta \mathcal{W}(i) \qquad (2)$$

where $\alpha_{nm}(i, h_{nm}(i))$ is the deterministic fading coefficient which can be predicted as described in *Stage I - Broadcasting*, and $\sigma_{nm}(i)\Delta \mathcal{W}(i)$ denotes the Wiener process that follows $N(0, \sigma_{nm}(i)\Delta t)$ for modeling the channel prediction uncertainty due to the small-scale fading. The initial channel value $h_{nm}(0)$ for all device $n$ and beam $m$ can be estimated from the downlink broadcast reference signal according to the reciprocity between TDD uplink and downlink channels[34]. Based on the initial channel value and Eq.(2), the channel states of the device $n$ at each frame can be derived.

### 3.2. Energy Evolution

Considering the limited battery capacity of IoT devices, the energy budget of each device within duration $T$ is assumed as $E_0$. The energy states evolution of device $n$ can be expressed as:

$$E_n(i+1) = E_n(i) - p_n(i, D_n(i))\Delta\tau, \quad E_n(I_{index}) \geq 0$$
$$= E_n(i) - \frac{p_n(t, D_n(i))\Delta t}{K} \quad (3)$$

in which $E_n(i)$ is the remaining energy at the end of the frame $i$, $p_n(i, D_n(i))$ represents the transmission power of the device $n$.

### 3.3. Transmission Model

With beamforming, the signal received at the BS can be expressed as:

$$y(i,D) = \underbrace{w_{nm}^H(i)h_{nm}(i)\sqrt{l_n \cdot p_n(i,D)}S_n(i,D)}_{\text{desire signal}} + \underbrace{\sum_{n' \in \Phi_m(i,D)/n} w_{n'm}^H(i)h_{n'm}(i)\sqrt{l_{n'} \cdot p_{n'}(i,D)}S_{n'}(i,D)}_{\text{intra-beam interference}} +$$
$$\underbrace{\sum_{m' \in \mathcal{M}/m}\sum_{n' \in \Phi_{m'}(i,D)} w_{n'm'}^H(i)h_{n'm'}(i)\sqrt{l_{n'} \cdot p_{n'}(i,D)}S_{n'}(i,D)}_{\text{inter-beam interference}} + w_{nm}^H(i)n_0$$

(4)

where $\Phi_m(i,D)$ is a subset of $N$ which selecting backoff delay $D$ and beam $m$ at frame $i$. $w_{nm} \in C^{L \times 1}$ is the beamforming vector of cluster $m$ and $(\cdot)^H$ denote the conjugate transpose. $S_n(i,D)$ represents the transmission signal of device $n$ where $\mathbb{E}(|S_n(i,D)|^2) = 1$. Moreover, $n_0$ is the power density of white Gaussian noise. Assuming that the BS can estimate perfect uplink CSI, we employ ZF beamforming to eliminate the inter-beam interference [9]. The BF matrix satisfies $\hat{W}_m^H(i) = H_m^H(i)(H_m(i)H_m^H(t))^{-1}$, in which $H_m(i) = [h_{1m}(i),...,h_{|\Phi_m(i)|m}(i)]$ is the collective vector channel between the device in cluster $m$ and the BS, and then apply the BF vector $w_{nm}^H(i) = \frac{\hat{w}_{nm}^H(i)}{|\hat{w}_{nm}^H(i)|}$, in which $\hat{w}_{nm}^H(i)$ is the $n$-th column of $\hat{W}_m^H(i)$.

A MPA decoder is assumed to be employed at the BS for SCMA decoding, which allows parallel decoding for different uplink signals from each device with different SCMA patterns in the same resource block (RB) [8]. Therefore, for a specific device signal, the SCMA signals of other devices in the same beam and RB can be treated as interference. When device $n$ selecting backoff delay $D_n(i)$, its signal-to-interference-noise-plus-ratio (SINR) at the BS can be denoted as:

$$\gamma_n(i, D_n(i)) = \frac{|w_{nm}^H(i)h_{nm}(i)|^2 \cdot l_n \cdot p_n(t, D_n(i))}{I_n(i, D_n(i)) + |w_{nm}^H(i)n_0|^2 B} \tag{5}$$

in which $|\cdot|$ is the Euclidean norm. And $B$ is the channel bandwidth, respectively.

It should be noted that, a low received signal to interference plus noise ratio (SINR) will lead to compromised decoding quality and diminished precoding effectiveness for the adopted ZF receiver, which in turn results in interference among devices distributed across different beams. Therefore, each device needs to ensure that SINR of its received signal at the BS is greater than the pre-defined SINR threshold $\gamma_n(i)$ when determining its backoff delay $D_n(i)$, i.e.,

$$\gamma_n(i, D_n(i)) \geq \gamma_0, \quad i \in \mathcal{I}, n \in N \tag{6}$$

For the convenience of writing, we assume that $\hat{H}_{nm}(i) = |w_{nm}^H(i)h_{nm}(i)|$, which satisfies:

$$\hat{H}_{nm}(i+1) = \hat{H}_{nm}(i) + \delta_{nm}^a(i)|w_{nm}^H(i)\alpha_{nm}(i, h_{nm}(i))|\Delta t + \delta_{nm}^b(i)|w_{nm}^H(i)\sigma_{nm}(i)|\Delta \mathcal{W}(i) \tag{7}$$

in which, $\delta_{nm}^a(i)$ and $\delta_{nm}^b(i)$ are sign functions and satisfy $\delta_{nm}^a(i) = \mathbf{sgn}(\alpha_{nm}(i, h_{nm}(i)))$, $\delta_{nm}^b(i) = \mathbf{sgn}(\sigma_{nm}(i))$.

The interference $I_n(i, D_n(i))$ received by device $i$ is caused by other devices in the same cluster that accidentally choose the same backoff delay, which can be represented as:

$$I_n(i, D_n(i))) = \sum_{n' \in \Phi_m(i, D_n(i)), n' \neq n} p_{n'}(i, D_n(i))l_{n'} \cdot \hat{H}_{nm}^2(i) \tag{8}$$

By inverting (5), the minimum required power $p_n^{req}(i, D_n(i))$ is obtained as:

$$p_n^{req}(i, D_n(i)) = \frac{\gamma_0}{\hat{H}_{nm}^2(i) \cdot l_n}\left[I_n(t, D_n(i)) + |w_{nm}^H(i)n_0|^2 B\right] \tag{9}$$

To minimize the energy consumption while maintaining a transmission quality constraint, we select $p^{req}$ as the transmission power and assume that the maximum transmission power of the device in each TS is $p^{max}$. When the channel condition is too poor, it may cause $p^{req} > p^{max}$, then the data packet is dropped in the current frame, and the data transmission is resumed in the next frame. Therefore, the transmission power can be expressed as:

$$p_n(i, D_n(i)) = \begin{cases} p_n^{req}(i, D_n(i)), & p^{req} \leq p^{max} \\ 0, & p^{req} > p^{max} \end{cases} \quad (10)$$

## 4. Problem Formulation

In the investigated scenario, each device $n$ needs to select its optimal backoff delay $\mathbf{D}_n^* = \{D_n^*(1),...,D_n^*(i),...,D_n^*(I_{index})\}$ for transmission frame $i = \{1,...,I_{index}\}$ from a bounded action set $D_n^*(i) \in \{\Delta\tau, 2\Delta\tau,..., K\Delta\tau\}$. The backoff delay should be minimized to ensure the effectiveness of its task data, under the long-term energy budget constraint $E_n(0)$, and based on the dynamic evolution of its remaining energy state $E_n(i)$ and channel states $h_{nm}(i)$. Thus, we adopt a cost function with distinct convexity[36], such as:

$$C_n(i) = D_n^2(i) \quad (11)$$

To facilitate the optimization process, $D_n(i)$ can be relaxed to a continuous space, and the obtained optimal value can be converted back to discrete value by rounding. Therefore, the optimization problem of backoff decisions for device $n$ can be defined as:

$$\begin{aligned}
\mathbf{D}_n^* &= \arg\min_{\mathbf{D}_n} \mathbb{E}\left[\int_0^T C_n(t)dt\right] \\
s.t. & \\
C1: \quad & d\hat{H}_{nm}(t) = \delta_{nm}^a(i)|w_{nm}^H(i)\alpha_{nm}(i, h_{nm}(i))|dt + \delta_{nm}^b(i)|w_{nm}^H(i)\sigma_{nm}(i)|d\mathcal{W}(t), \\
C2: \quad & dE_n(t) = -\frac{p_n(t, D_n(t))}{K}dt, \\
C3: \quad & E_n(0) = E_n^0, \\
C4: \quad & h_{nm}(0) = h_{nm}^0, \\
C5: \quad & E_n(T) \geq 0.
\end{aligned} \quad (12)$$

in which $C1$ and $C2$ describe the evolution of the channel gain and the remaining energy state of device $n$, respectively; $C3$ and $C4$ represent the initial energy and channel states, respectively. Each device $n$ attempts to solve its own version of the optimization problem (12) at the same time, leading to an n-player non-cooperative dynamic stochastic game (DSG). Based on the dynamic programming theory [36], the optimal solution of (12) within duration $[0,T]$ is to solve the Bellman running cost function in a time-reversed order, which can be defined as:

$$v_n(t, X_n(t)) = \min_{\mathbf{D}_n} \mathbb{E}\left[\int_t^T C_n(u)du + F(E_n(T))\right] \quad (13)$$

where

$$X_n(t) = [E_n(t), \hat{H}_{nm}(t)] \tag{14}$$

is the state of device $n$ at time t, composed of the remaining energy $E_n(t)$ and the channel state $\hat{H}_{nm}(t)$. $F(E_n(T))$ represents the penalty function that penalizes the case of exhausting all energy before time $t$. If $E_n(T) \leq 0$, $F(E_n(T))$ should be an appropriately large positive value; if $E_n(T) \geq 0$, $F(E_n(T)) = 0$. In this work, a parametric logistic penalty function is adopted as:

$$F(E_n(T)) = \frac{\phi}{1+e^{\rho E_n(T)}} - \frac{\phi}{2}.$$

**Definition 1**: The optimal backoff strategy $\mathbf{D}^*(t) = \{D_1^*(t), ..., D_n^*(t), ..., D_N^*(t)\}$ is a Nash Equilibrium (NE) for the n-player DSG described in (12), if and only if $\mathbf{D}^*(t)$ is the optimal control for the following problem:

$$D_n^*(t) = \arg\min_{D_n^*(t)} \mathbb{E}\left[\int_t^T C_n(u, D_n(u), D_{-n}^*) du + F(E_n(T))\right] \tag{15}$$

where $D_{-n}^*$ represents the backoff strategies of all the devices except device $n$. Under the NE definition, none of the devices can achieve lower cost by deviating from its optimal backoff strategy unilaterally.

Based on [37] the sufficient condition for the existence of the NE is that the running cost function $v_n(t, X_n(t))$ for $n$ devices has a solution to its HJB equation, which can be guaranteed by the smoothness of the Hamiltonian $H_{am}$. In this optimization problem, the HJB equation and the corresponding Hamiltonian for each device are shown in Eq. (16) and Eq. (17), respectively

$$\partial_t v_n^*(t, X_n(t)) + \min_{D_n(t)}\left[-(\frac{\gamma_0}{K \cdot l_n \cdot \hat{H}_{nm}^2(t)}[I_n(t, D_n(t)) + |w_{nm}^H(t)n_0|^2 B])\partial_{E_n(t)} v_n^* \right.$$
$$\left. +\delta_{nm}^a(t)|w_{nm}^H(t)\alpha(t, h_n(t))|\partial_{\hat{H}} v_n^* + \frac{\delta_{nm}^b(t)|w_{nm}^H(t)\sigma_{h,n}|^2}{2}\partial_{\hat{H}\hat{H}}^2 v_n^* + D_n^2(t)\right] = 0 \tag{16}$$

$$H_{am}(D_n(t), X_n(t)) = \min_{D_n(t)}\left[-(\frac{\gamma_0}{K \cdot l_n \cdot \hat{H}_{nm}^2(t)}[I_n(t, D_n(t)) + |w_{nm}^H(t)n_0|^2 B])\partial_{E_n(t)} v_n^* \right.$$
$$\left. +\delta_{nm}^a(t)|w_{nm}^H(t)\alpha(t, h_n(t))|\partial_{\hat{H}} v_n^* + \frac{\delta_{nm}^b(t)|w_{nm}^H(t)\sigma_{h,n}|^2}{2}\partial_{\hat{H}\hat{H}}^2 v_n^* + D_n^2(t)\right] \tag{17}$$

Proof: See Appendix A.

To obtain the optimal control strategy $D_n(t)$, given that this is a convex optimization problem, we take the partial derivative of the function and set it to zero, resulting in Eq. (18):

$$D_n^*(t) = \frac{\gamma_0}{2K \cdot l_n \cdot \hat{H}_{nm}^2(t)} \frac{\partial I(t, D_n(t))}{\partial D_n(t)} \frac{\partial v_n^*(t, X_n(t))}{\partial E_n(t)} \qquad (18)$$

Proof: See Appendix B.

According to the proof in Appendix B, the Hamiltonian is smooth, which implies the existence of the Nash equilibrium [38]. However, it must be noted that in Eq. (18), the interference term $I(t, D_n(t))$ represents the cumulated result of $D_{-n}^*$ for $D_n^*(t)$ of device $n$, which means that $n$ correlated partial differential equations (PDEs) need to be solved simultaneously. As $n$ becomes large, this task would become prohibitively difficult. To address this scaling problem, we next transform the problem into a MFG, which provides better tractability.

## 5. Mean Field Game Approach

In this section, MFG [39] is introduced to convert the n-player non-cooperative game into the interaction between only two bodies, namely the generic device and the mean field, such that the problem can be solved no matter how large is n. Then the MFE is derived with both HJB and FPK equations, and the corresponding Mean Field-based Dynamic Backoff (MFDB) algorithm is proposed.

### 5.1. Problem Reformulation with Mean Field Theory

According to the mean field theory [22], a MFG model consists of a generic player who takes rational actions and a mean field representing the collective actions of all other players. When the game starts, the generic player devises a decision set for all possible states to optimize their cost, which is shared among all players. Subsequently, the mean field, using its probability density function (PDF), calculates the cumulative impact of all other players on the generic player based on this shared decision set. In response, the generic player adjusts their decisions based on the mean field's feedback. The mean field then updates its impacts reflecting the new decision set. This iterative process continues until a NE is achieved. It is obvious that in a MFG, which functions as a two-body game, the convergence time does not increase with the number of players.

In a MFG framework [23], the model features a typical agent who follows rational decision-making, and a mean field that aggregate the behavior of all other agents who are also rational. As the game commences, this typical agent formulates a strategy for all conceivable states to minimize its associated cost, which is uniformly adopted by all the agents in the game. Then the PDF of the mean field can be employed to calculate the collective effect of all the typical agents, leveraging the common strategic framework. In reaction to the impact of the mean field, the typical agent fine-tunes its strategy accordingly. The mean field, in turn, updates its effects to reflect the revised strategy. This dynamic interaction will continuous until a NE is reached. It is obvious that the convergence time of a MFG, which essentially operates as a two-agent interaction, remains stable regardless of the number of agents.

To formulate a MFG, four hypotheses need to be satisfied:

- *H1 - A continuum of a large number of players:* Assuming a sufficiently large number of IoT devices participating in the game, such that it can be approximated as infinite. Since the number of clusters is limited and far smaller than the number of devices, the number of devices in each cluster can also be considered infinite so that the devices can be regarded as the player continuum.
- *H2 - The player's rational behaviors:* It is assumed that the devices involved in the game have rational behavior. The devices will all implement the optimal backoff delay at any given time, and it will depend exclusively on the current state $X_n(t)$ they are in, which makes these strategies predictable for other devices.
- *H3 - The interchangeability of the players*: Since the optimal backoff strategy of each device only depends on its state and the interference of other devices. Therefore, changing the order of devices does not change their backoff decision. Devices in the same state will have the same backoff delay. Based on this assumption, we can decide the backoff delay based on the state of the device rather than n separate strategies.
- *H4 - The mean field can describe the interaction between players*: For a single device *n*, instead of considering the one-to-one interaction, we only consider the jointly affected by $\Phi_m(t, D_n(t)) - 1$ other devices, namely the intra-beam interference, which consists of the weighted sum of the transmission power of other devices in the same cluster under the same backoff delay. Due to the above three characteristics, we can convert the interference into the mean field interference based on the backoff delay strategy and the distribution of system states.

Given the investigated system satisfies *H1-H4*, the DSG problem (12) can be transformed to a MFG as follows:

**Definition 2**: For the state space $X_n(t) = [E_n(t), \hat{H}_{nm}(t)]$, the mean field is the probability distribution of this state space at time *t*, where the PDF of users in any specific state is:

$$m(t, X) = \lim_{N \to \infty} M(t, X) = \lim_{N \to \infty} \frac{1}{N} \sum_{n=1}^{N} 1_{X_n(t) = X} \tag{19}$$

in which $M(t, X)$ represents the proportion of devices in state X at frame *t*. 1 denotes the indicator function that returns 1 when the given condition is satisfied, otherwise it returns 0. The density function $M(t, X)$ will converge to the mean field density $m(t, X)$ as the number of devices n tends to infinity which satisfies:

$$\int_{H \in \mathcal{H}} \int_{E \in \mathcal{E}} m(t, X) dh dE = 1 \tag{20}$$

in which $\mathcal{H}$ and $\mathcal{E}$ are the set of channel gain and remaining energy of all devices, respectively. $m(t, X)$ is a continuous PDF. The optimal backoff delay can be determined by solving the HJB equation. We denote the proportion of devices with the same backoff delay $D$ at frame $t$ and the corresponding device state distribution by $\Lambda(t, D)$ and $G(t, D, X)$, respectively. As $n$ tends to infinity, they can be converted into $\lambda(t, D)$ and $g(t, D, X)$, which are continuous PDFs and can be deduced as:

$$\lambda(t, D) = \lim_{N \to \infty} \Lambda(t, D) = \lim_{N \to \infty} \frac{1}{N} \sum_{n=1}^{N} 1_{D_n(t, X) = D} \tag{21}$$

$$g(t, D, X) = \lim_{N \to \infty} G(t, D, X) = m(t, X) \cdot \lambda(t, D) \tag{22}$$

They also satisfy the following conditions:

$$\int_{D \in \mathcal{D}} \lambda(t, D) dD = 1 \tag{23}$$

$$\int_{H \in \mathcal{H}} \int_{E \in \mathcal{E}} \int_{D \in \mathcal{D}} g(t, D, X) dH dE = 1 \tag{24}$$

in which, $\mathcal{D}$ is the set of the backoff delay that the device can choose. Therefore, the number of the devices that select the same TS to transmit in the same cluster $|\Phi_m(t, D)|$ can be expressed as:

$$|\Phi_m(t, D)| = |\Phi_m(t)| \cdot \lambda(t, D) \tag{25}$$

$$\partial_t v^*(t, X) + \min_{D(t)} \left[ -\frac{\gamma_0}{K \cdot \hat{H}_m^2(t)} \left( I(t, D) + |w_m^H(t) n_0|^2 B \right) \partial_{E(t)} v^*(t, X) \right.$$
$$\left. + \delta_m^a(t) |w_m^H(t) \alpha(t, h_m(t))| \partial_{\hat{H}} v^*(t, X) + \frac{\delta_m^b(t) |w_m^H(t) \sigma_h|^2}{2} \cdot \partial_{\hat{H}\hat{H}}^2 v^*(t, X) + D^2(t, X) \right] = 0 \tag{26}$$

Due to the fixed device density $\rho$, the interference term in (8) will converge to a constant value that depends on the device density as the number of devices increases [40]. In order to describe the interaction between devices with the mean field, we transform the interference into the mean field interference and guarantee its boundedness. That is Eq.9 is rewritten as:

$$I_n(t) = \frac{\beta}{|\Phi_m(t)| - 1} \sum_{n' \in \Phi_m(t, D_n(t)), n' \neq n} p_{n'}(t, D_n(t)) \hat{H}_{n'm}^2(t) \tag{27}$$

where $\beta$ denotes the normalized interference factor depending on the path loss index and the device density.

The complete proof is presented in Appendix C.

Then, the interference term can be converted from (27) to:

$$I_n(t) = \frac{\beta \cdot |\Phi_m(t, D_n(t))|}{|\Phi_m(t)|-1} \sum_{H \in \hat{\mathcal{H}}} \sum_{E \in \mathcal{E}} p(t, D_n(t), X) \cdot m(t, X) \hat{H}_m^2(t) - p_n(t, D_n(t)) \hat{H}_{nm}^2(t) \quad (28)$$
$$= \frac{\beta \cdot |\Phi_m(t)| \lambda(t, D_n(t))}{|\Phi_m(t)|-1} \sum_{H \in \hat{\mathcal{H}}} \sum_{E \in \mathcal{E}} p(t, D_n(t), X) \cdot m(t, X) \hat{H}_m^2(t) - p_n(t, D_n(t)) \hat{H}_{nm}^2(t)$$

When $|\Phi_m(t)|$ approaches infinity, according to H1-H4, all the devices will rationally select the optimal backoff $D_n(t) = D_n^*(t)$ (from Eq.(18)), therefore (28) can be calculated based on the continuous mean-field PDF in (21) and (22), such as:

$$I_n(t) = \beta \lambda(t, D_n^*) \cdot \int_{H \in \hat{\mathcal{H}}} \int_{E \in \mathcal{E}} m(t, X) \cdot p(t, D_n^*, X) \hat{H}^2(t) dH dE \quad (29)$$

From (28) and (29), it can be seen that devices transmitting with the same backoff delay will approximately suffer the same cumulative interference as the number of devices tends to infinity. Therefore, we can ignore the device index n and establish the relationship between backoff delay and interference:

$$I(t, D) = \beta \lambda(t, D) \cdot \int_{H \in \hat{\mathcal{H}}} \int_{E \in \mathcal{E}} m(t, X) \cdot p(t, D, X) \hat{H}^2(t) dE dH \quad (30)$$

**5.2. Mean Field-based Dynamic Backoff Scheme**

To this end, the N-body problem in (12) can be converted to an equivalent MFG, viewed as a two-body problem, as illustrated in Fig. 3. Then we explain how the optimal control $D^*$ to achieve the MFE will be derived from the interaction between these two bodies.

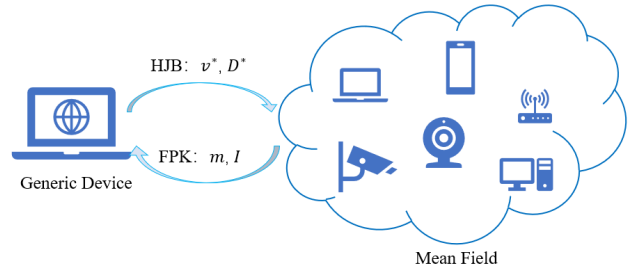

**Figure 3:** Graphical explanation of applying two body MFG to N-body computational backup decisions

- First body - Generic Device: According to the HJB equation, each device can decide its optimal backoff delay based on its state. The general HJB equation is expressed as (26) at the bottom of the page, and the index *n* in (18) can be removed, leading to the optimal backoff policy for a generic device as follow:

$$D^*(t) = \frac{\gamma_0}{2K \cdot l \cdot \hat{H}_m^2(t)} \frac{\partial I(t, D(t))}{\partial D(t)} \frac{\partial v(t, X(t))}{\partial E(t)} \quad (31)$$

- Second body - Mean Field: The cumulated interference to a generic device is now sufficiently described by (30), in which the evolution of the mean field PDF can be derived as [22]:

$$\partial_t m(t,X) + \partial_H (\delta_m^a(t) | w_m^H(t)\alpha(t) | m(t,X)) - p(t,D^*(t),X)\partial_E m(t,X) - \frac{1}{2}\delta_m^b(t) | w_m^H \sigma |^2 \partial_{hh} m(t,X) = 0 \quad (32)$$

Proof: See Appendix D.

As presented in Fig. 3, the HJB equation (26) is employed to derive the optimal backoff strategy (31) to be used for any device in any states (channel state, remaining energy) under the initial mean field interference (from any initial mean field PDF), while the FPK equation (32) allows to calculate the mean field interference (30) given all devices in the system follow the optimal backoff strategies from HJB since they are all rational. After that, the HJB will recalculate the optimal control solution according to the updated mean field interference, then the FPK will derive the new mean field evolution based on the updated backoff control. This interactive process will be repeated until the optimal control or its corresponding value function converge, as shown in **Algorithm 1**.

---

**Algorithm 1** The Proposed Mean Field-based Dynamic Backoff algorithm (MFDB)
---
1: **Initialize:** $v_{(0)}^*, m_{(0)}^*, I_{(0)}^*, D_{(0)}^*, f=0, f_{max}, \varepsilon$
2: **While** $f < f_{max}$ **do**
3:      $f = f+1;$
4:      Calculate (26) to obtain the value of $v_{(f)}^*$ based on $D_{(f-1)}^*$.
5:      Calculate (31) to update the value of $D_{(f)}^*$ based on $V_{(f)}^*$ and $I_{(f-1)}^*$.
6:      **if** $|D_{(f)}^* - D_{(f-1)}^*| < \varepsilon$ **then**
7:          break;
8:      **end if**
9:      Update the value of $p_{(f)}^*$ based on $D_{(f)}^*$ and $I_{(f-1)}^*$.
10:     Calculate (32) to obtain the value of $m_{(f)}^*$ based on $p_{(f)}^*$.
11:     Update the value of $I_{(f)}^*$ based on $D_{(f)}^*$ and $m_{(f)}^*$.
12: **end while**

---

### 5.3. Optimality Analysis

In the MFG, when the individual strategies (their optimal policy in (31)) and the mean field reach a stable state, where no device can increase its value by unilaterally changing their strategy, the system reaches a Mean Field Equilibrium (MFE), which can be seen as the equivalent to the Nash equilibrium for the *n*-player DSG in (16) before MFG is employed. At this point, each device's strategy is the best response to the strategies of all others. In our system, at any time $t$ and state $X$, the value function $v(t,X)$ and the mean field $m(t,X)$ interact with each other, where the optimal value $v^*(t,X)$ is determined by solving the HJB equation, as described in (31), and $m(t,X)$ is the solution to the FPK equation in (32). The optimal value $v^*(t,X)$ determines the optimal strategy $D^*(t,X)$ in (31), which influences the evolution of the mean field $m(t,X)$ via (32). This, in turn, determines the mean interference $I_n(t)$ in (30), which affects $v^*(t,X)$ through (26). Therefore, the optimal strategy can be obtained by iteratively solving two coupled forward-backward PDEs. Since all the functions involved are smooth, the iterative algorithm is guaranteed to converge to the optimal mean field strategy [42], thereby bringing the system into the MFE state.

**5.4. Complexity Analysis**

The computational complexity of the *n*-player DSG in Sec.4 and the proposed MFG-based **Algorithm 1** are compared as follows.

- N-player DSG: In the model, each device is required to account for both its own action and the actions of all other devices, by solving (16)-(18) for N devices at the same time. This integration leads to a significant increase in action space and computational complexity as the number of devices *N* grows, e.g., if the action space of each device is *A*-dimensional, then the total action space of the system becomes $A^N$.
- MFG: The MFG simplifies the interactions between *N* devices by transforming the complex multi-player game into a two-player game, where each individual interacts with the average behavior of all the others. In other words, the mean field simplifies the complex interactions of a large number of participants into interactions between individuals and the mean field. This method introduces the mean field approximation, which transforms the high-dimensional game problem of n devices into a game between an individual and the mean field of the overall system. This approach substantially reduces the complexity by limiting the system action space to $A^2$, thus significantly reducing the computational complexity. Consequently, the MFG-based **Algorithm 1** will converge fast, since its HJB-FPK iterations only involve two players instead of the n players in the N-body DSG [42].

**6. Numerical Results**

In this section, we employ the FDM to solve the proposed MFDB scheme numerically, as described in **Algorithm 1**. Since ZF precoding eliminates the inter-beam interference, all of the following numerical results are for the device in one spatial beam, and the devices in other beams follow the same strategy. To maintain generality and ensure consistency, the system states $E$ can be normalized to the interval [0, 1]. Table 1 presents the key simulation parameters employed in our work.

**Table 1:** Simulation Parameters

| parameter | value |
|---|---|
| Number of frames $I_{index}$ | 20 |
| Frame duration $\Delta t$ | 10 $ms$ |
| Number of TSs per frame $K$ | 20 |
| TS duration $\Delta \tau$ | 0.5 $ms$ |
| Path loss exponent $a$ | 2.5 |
| Noise power $N_0$ | -104 $dBm$ |
| Channel bandwidth $B$ | 50 $kHz$ |
| Device density $\rho$ | 5 per square meter |
| SINR threshold $\gamma_0$ | 0.5 |
| Number of devices $N$ | 1000 |

### 6.1. Semi-static Channels

We assume a semi-static channel with constant channel gain during the simulations in Fig. 4 and Fig. 5. Fig. 4 describes the optimal backoff decisions $D_n^*(t, E)$ for each device with a constant channel gain $H_c = \hat{H}_0^2 \cdot l = 3 \times 10^{-3}$, which reveals that the backoff delay for a specific frame decreases as the remaining energy increases. Moreover, when the remaining energy is fixed, the IoT devices can adopt a lower backoff delay as the frame index gets closer to the final one. This is due to the fact that devices with sufficient energy will reduce the worry of running out of energy budgets and incurring penalties before the transmission deadline. Fig. 5 shows the evolution of the optimal mean field distribution $m_n^*(t, E)$ where the initial mean field $m(0, E)$ is uniformly distributed in all energy states under the constant channel gain $H_c$. The figure reveals that most devices just run out of energy or have energy left by the end of the transmission duration. Only a few devices experience penalties due to insufficient initial energy to complete the data transmission leading to an early exhaustion of energy.

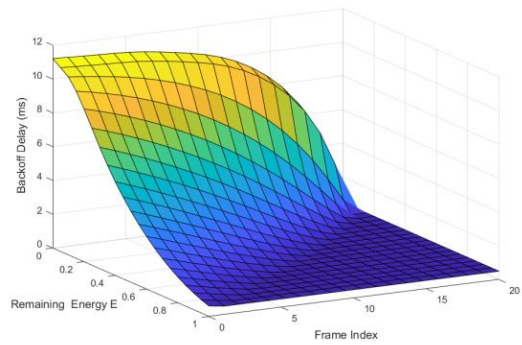

**Figure 4:** The Optimal backoff delay *D* under the constant channel.

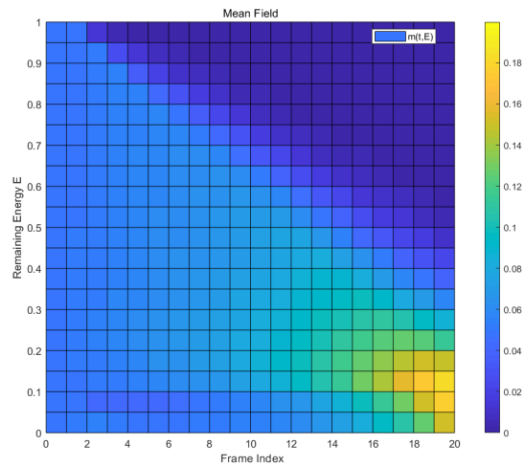

**Figure 5:** Mean field evolution under a constant channel gain

### 6.2. Dynamic Channels

As in Eq. (7), the dynamic channel evolution is modeled as a stochastic differential equation with the uncertainty coefficient $\sigma$. In Fig. 6, we evaluate the impact of channel uncertainty on the backoff delay of MFDB by considering the following:

- h1: The certain channel with $\sigma = 0$.
- h2: The low unpredictable channel with $\sigma = 0.1$.
- h3: The medium unpredictable channel with $\sigma = 1$.
- h4: The high unpredictable channel with $\sigma = 10$.

All the above channel scenarios have the same deterministic part, i.e.,

$$\hat{H}_d(t) = H_c + A\sin(f_0 t + \theta) \tag{33}$$

where $H_c = 3 \times 10^{-3}$, $A = 2 \times 10^{-3}$, $f_0 = 0.4$, $\theta = 2$. It can be observed in Fig. 6(a) as the channel uncertainty $\sigma$ gets larger, the uncertainty of the channel increases. This indicates that the channel quality deviates more from the predicted channel evolution. In Fig. 6(b), we depict the effect of channel uncertainty on the backoff delay in the MFDB strategy. It can be seen that compared with the fully predictable channel h1, the higher the uncertainty of the channel, the higher the backoff delay will be selected by the MFDB strategy. This is because when the channel becomes highly unpredictable, the device cannot judge whether the remaining energy can support the data transmission. In this case, the device may not accurately estimate the energy required for data transmission at a certain moment, which will increase the risk of transmission failure and waste precious energy resources. This strategy helps to maintain the availability of the device in the long term and avoid the device stopping working due to energy exhaustion.

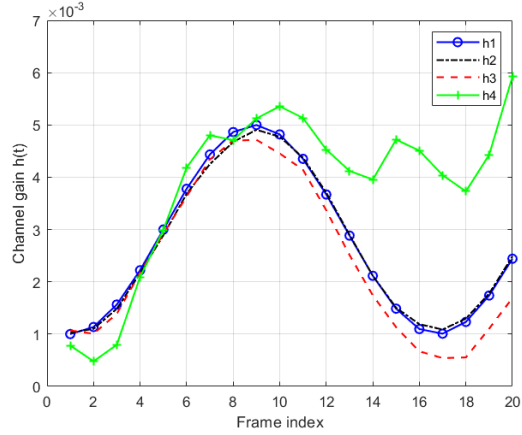

(a) Predicted channel evolution

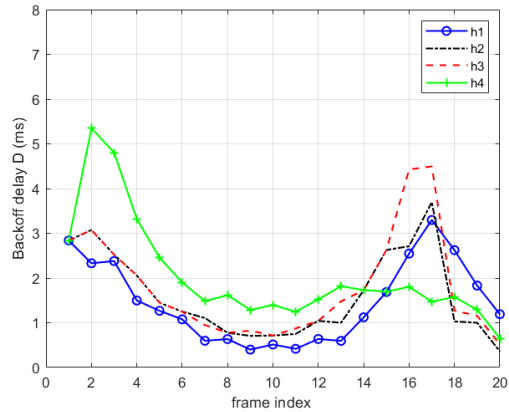

(b) Optimal backoff delay

**Figure 6:** The optimal backoff delay D under different stochastic channels for a generic device

### 6.3. Comparison with other backoff schemes

In this subsection, we compare the performance of the MFDB scheme with other backoff schemes, which are:

- ACB: The BS generates an ACB factor $b_0$ in each frame and broadcasts it to the device. Then, the device generates a random number $b \in [0,1]$ before sending the data and compares it with the ACB factor. If $b > b_0$, the device transmits data with a fixed transmission power using SCMA. The base station determines whether the decoding is successful according to the received SINR of a specific device. Then the base station sends a feedback ACK or NACK signal ("ACK" for success while "NACK" for failure) to inform the device. If receiving a NACK, the device will randomly backoff for 1 to 3 TSs and resend the data packet.

- Slotted-ALOHA: The device randomly selects a backoff delay in each frame to transmit data with a fixed transmission power using SCMA. The base station determines whether the decoding is successful according to the received SINR and sends a feedback signal with ACK or NACK. If the decoding fails, the device will randomly backoff for 1 to 3 TSs and retransmit the data.
- Minimum backoff (MB): In this baseline scheme, the device will always transmit SCMA data in each frame's first TS. The interference is determined by all device power and channel state, which is pre-counted by the BS and broadcast to all devices in each frame [27]. The device decides the transmission power based on the interference level.

To evaluate the backoff delay of the above scheme, we consider the following channel scenarios:

- *Constant channel (CC)*: The channel gain is consistently $h_0$.

- *Dynamic channel (DC)*:   According to (7), the DC is modeled as two parts, where the deterministic part follows (36) with different parameter $f_0 = 20$ and variance $\sigma = 0.1$.

The normalized energy budget $E(0) = 0.7$ in the following simulations, and the simulated device number is 1000. As shown in Fig. 7, for ACB and slotted-ALOHA scheme, the backoff delay is the average result of all the devices due to the random backoff. For the MFDB scheme, since all devices follow the same backoff strategy, the figure depicts the expected result for a generic device. Fig. 7 reveals that whether it is under CC or DC, the MB scheme cannot complete the data transmission in all frames. This is because when all the devices are transmitting with the minimum backoff, high transmitting power will be required to overcome the severe interference among devices.  Therefore, the remaining energy in this scheme is used up before the end of the transmission, regardless of the channel condition. For the MFDB scheme, the backoff delay remains relatively constant under CC, even if the device's energy decreases throughout the frame evolution. This is due to the fact that the device is able to predict the continuous decrease of the other devices' energy with the mean field. And the device is able to dynamically adjust its backoff delay according to the changing channel under DC.

Moreover, it can be seen from the figure that the MFDB significantly outperforms the ACB and the slotted-ALOHA scheme in terms of backoff delay. That is because the MFDB scheme can dynamically adjust the backoff in each frame according to its current channel gain and remaining energy. Thus, the MFDB scheme can avoid the case that a large number of devices access the same TS resulting in high decoding failure probability at the BS and extra delay due to data re-transmissions.

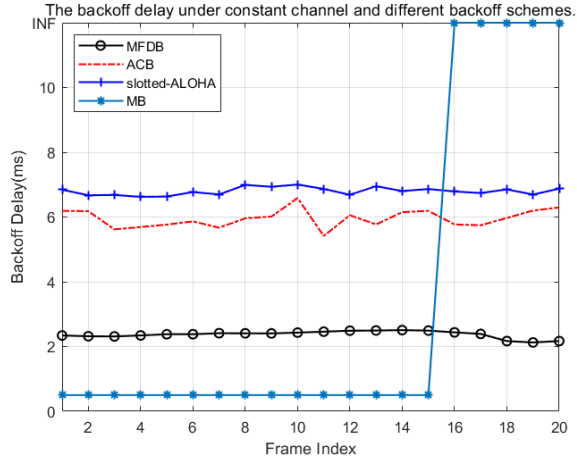

(a) Under CC

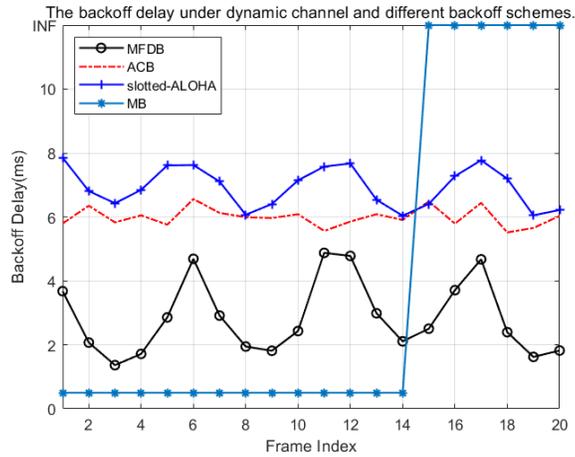

(b) Under DC

**Figure 7:** The backoff delay D under different backoff scheme for a generic device. (a) Under CC; (b) Under DC

Fig. 8 depicts the cumulated delay cost (CDC) for the four evaluated strategies. According to (11), $CDC(t)$ is defined as $CDC(t) = \sum_{i=1}^{t} D_n^2(i)$. It can be observed that the CDC of the MFDB scheme demonstrates a significantly slower increase compared to the ACB and slotted-ALOHA schemes, regardless of the channel condition being CC or DC. However, because the MB scheme is transmitted continuously on the first TS of each frame, the remaining energy is used up early. When the remaining energy is exhausted, the device cannot transmit, and the corresponding CDC is denoted as INF.

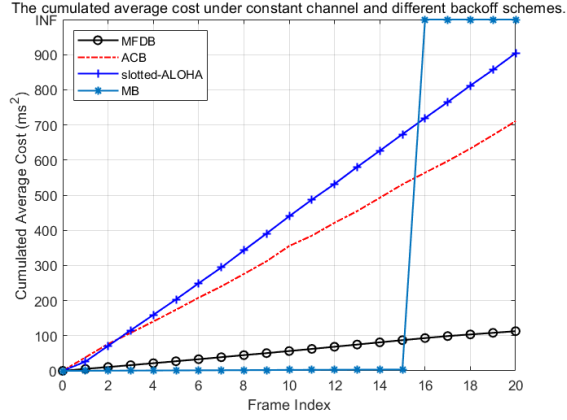

(a) Under CC

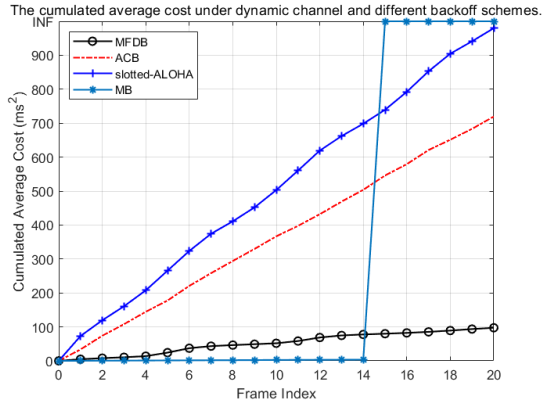

(b) Under DC

**Figure 8:** The CDC under different backoff scheme for a generic device (a) Under CC; (b) Under DC

Fig. 9 illustrates the average backoff delay versus the number of devices in the beam with different backoff strategies under different channel conditions. It can be observed that no matter what channel condition, the device with MFDB strategy always maintains the lowest backoff delay, which has little growth trend and is almost independent of the number of devices. When the number of devices is less than 900, the backoff delay of ACB and Slotted-ALOHA tends to be stable, and the backoff delay of ACB is slightly higher than that of slotted-ALOHA. This is because the average number of transmitting devices per slot in this case is less than the threshold for the number of devices that can be successfully decoded. Moreover, the random factor judgment of the ACB strategy will increase the backoff delay. When the number of devices is between 900 and 1300, the backoff delay of Slotted-ALOHA rapidly exceeds that of ACB. This is due to the increased probability of decoding failure in this case, and the random factor judgment of ACB can adjust the number of access devices to reduce the probability of decoding failure. When the number of devices exceeds 1300, in this case, the random factor of ACB also fails to alleviate the decoding failure but increases the backoff delay.

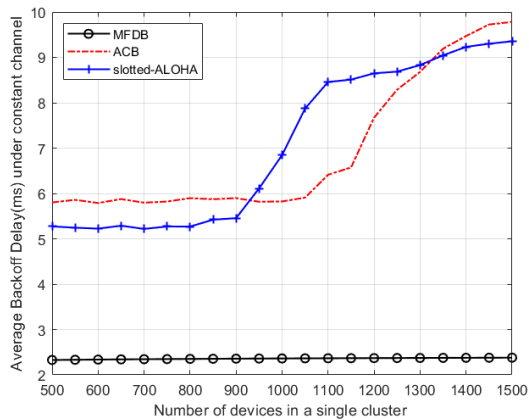

(a) Under CC

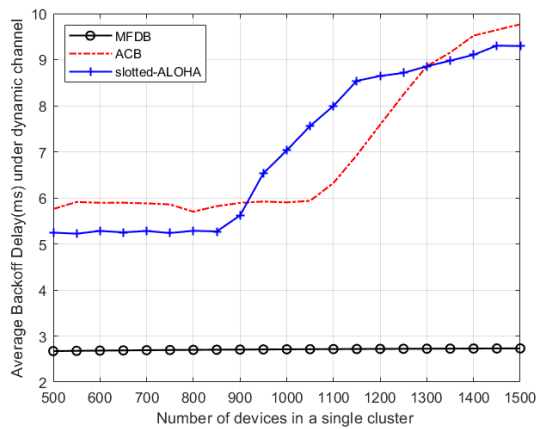

(b) Under DC

**Figure 9:** The average backoff delay versus device number under different channel condition. (a) Under CC; (b) Under DC

## 7. Conclusion and Future Work

In this work, we investigate the optimal dynamic backoff mechanism for massive random access within a 6G ultra-dense IoT system. Considering a 6G cell employing GF-NOMA and multi-beam MIMO, we design a clustering scheme based on GoB and an access signaling process based on GFRA. A MFDB scheme is proposed for each cluster to minimize the long-term cost of backoff delay of a generic device. Numerical results validate that the proposed MFDB can proactively adjust the backoff delay and transmission power according to the predicted channel gain and energy level evolution subject to the specified energy constraints. Compared with three other GFRA schemes, namely ACB, Slotted-ALOHA, and MB, the proposed MFDB mechanism can significantly reduce the average access delay and maintain a nearly constant backoff delay level even as the number of active devices achieves 2000 in a single subcarrier per cell.

In future work, we intend to setup real-world experiment environment to implement the proposed

MFDB scheme and to evaluate its validity. Meanwhile, we would also add other evaluation indicators such as energy efficiency to evaluate the performance of our proposed method. After that, the proposed MFG approach needs to be extended to multi-cell and multi-channel cellular systems with combined backoff delay, frequency resource, and NOMA preamble selections.

**Acknowledgement:** This work was supported by the National Natural Science Foundation of China.

**Funding Statement:** This work was supported by the National Natural Science Foundation of China under Grant 62371036, supported author H. Wang, H. Gao and P. Jiang. Website: https://www.nsfc.gov.cn/english/site_1/index.html. [Accessed: July 25, 2024]

**Author Contributions:** Haibo Wang: Provided the problem formulation, proposed the idea of Mean-field Game -based backoff scheme, and revise the JIoT manuscript for many times; Hongwei Gao: contributed the major writing of the journal paper, and most of the math derivation and simulations. Pai Jiang: wrote the related conference paper, and made the basic simulation for the conference paper. During the writing and submission of the JIoT paper, she had graduated from her master study and cannot make further contribution to the journal paper. Matthieu De Mari: who had co-supervise both Pai Jiang and Hongwei Gao in deriving the MFG solutions. Panzer Gu: give guidance on how to adopt the MFG backoff strategy in the MIMO-enabled cellular IoT systems, and how to design the grant-free access procedure. Yinsheng Liu: give guidance on how to design the MIMO channel model and express it in partial differential equations.

**Availability of Data and Materials:** Data available on request from the authors. "The data that support the findings of this study are available from the corresponding author, [Gao], upon reasonable request."

**Conflicts of Interest:** We declare that they have no conflicts of interest to report regarding the present study".

## References

[1] P. Jiang, H. Wang and M. De Mari, "Optimal Dynamic Backoff for Grant-Free NOMA IoT Networks: a Mean Field Game Approach," 2022 IEEE/CIC International Conference on Communications in China (ICCC), Sanshui, Foshan, China, 2022, pp. 997-1002

[2] M. Dohler, and S. J. Johnson, "Massive Non-Orthogonal Multiple Access for Cellular IoT: Potentials and Limitations," in IEEE Communications Magazine, vol. 55, no. 9, pp. 55-61, Sept. 2017.

[3] Y. Liu, Y. Deng, M. Elkashlan, A. Nallanathan and G. K. Karagiannidis, "Optimization of Grant-Free NOMA With Multiple Configured-Grants for mURLLC," in IEEE Journal on Selected Areas in Communications, vol. 40, no. 4, pp. 1222-1236, April 2022.

[4] J. Choi, J. Ding, N. -P. Le and Z. Ding, "Grant-Free Random Access in Machine-Type Communication: Approaches and Challenges," in IEEE Wireless Communications, vol. 29, no. 1, pp. 151-158, February 2022.

[5] J. Zhang, X. Tao, H. Wu, N. Zhang and X. Zhang, "Deep Reinforcement Learning for Throughput Improvement of the Uplink Grant-Free NOMA System," in IEEE Internet of Things Journal, vol. 7, no. 7, pp. 6369-6379, July 2020.

[6] M. Fayaz, W. Yi, Y. Liu and A. Nallanathan, "Transmit Power Pool Design for Grant-Free NOMA-IoT Networks via Deep Reinforcement Learning," in IEEE Transactions on Wireless Communications, vol. 20, no. 11, pp. 7626-7641, Nov. 2021.

[7] J. Liu, G. Wu, X. Zhang, S. Fang and S. Li, "Modeling, Analysis, and Optimization of Grant-Free NOMA in Massive MTC via Stochastic Geometry," in IEEE Internet of Things Journal, vol. 8, no. 6, pp. 4389-4402, 15 March15, 2021.

[8] B. Wang, K. Wang, Z. Lu, T. Xie and J. Quan, "Comparison study of non-orthogonal multiple access schemes for 5G," 2015 IEEE International Symposium on Broadband Multimedia Systems and Broadcasting, Ghent, Belgium, 2015, pp. 1-5.

[9] W. Yuan, N. Wu, Q. Guo, Y. Li, C. Xing and J. Kuang, "Iterative Receivers for Downlink MIMO-SCMA: Message Passing and Distributed Cooperative Detection," in IEEE Transactions on Wireless Communications, vol. 17, no. 5, pp. 3444-3458, May 2018.

[10] A. Almradi, P. Xiao and K. A. Hamdi, "Hop-by-Hop ZF Beamforming for MIMO Full-Duplex Relaying With Co-Channel Interference," in IEEE Transactions on Communications, vol. 66, no. 12, pp. 6135-6149, Dec. 2018.

[11] W. A. Al-Hussaibi and F. H. Ali, "Efficient User Clustering, Receive Antenna Selection, and Power Allocation Algorithms for Massive MIMO-NOMA Systems," in IEEE Access, vol. 7, pp. 31865-31882, 2019.

[12] S. Gong, C. Xing, V. K. N. Lau, S. Chen and L. Hanzo, "Majorization-Minimization Aided Hybrid Transceivers for MIMO Interference Channels," in IEEE Transactions on Signal Processing, vol. 68, pp. 4903-4918, 2020.

[13] X. Ge, W. Shen, C. Xing, L. Zhao and J. An, "Training Beam Design for Channel Estimation in Hybrid mmWave MIMO Systems," in IEEE Transactions on Wireless Communications, vol. 21, no. 9, pp. 7121-7134, Sept. 2022.

[14] S. Duan, V. Shah-Mansouri, Z. Wang and V. W. S. Wong, "D-ACB: Adaptive Congestion Control Algorithm for Bursty M2M Traffic in LTE Networks," in IEEE Transactions on Vehicular Technology, vol. 65, no. 12, pp. 9847-9861, Dec. 2016.

[15] T. Tao, F. Han and Y. Liu, "Enhanced LBT algorithm for LTE-LAA in unlicensed band," 2015 IEEE 26th Annual International Symposium on Personal, Indoor, and Mobile Radio Communications (PIMRC), 2015, pp. 1907-1911.

[16] M. R. Amini, A. Al-Habashna, G. Wainer and G. Boudreau, "Performance Analysis of Random Access NOMA for Critical mIoT With Timer-Power Back-Off Strategy," in IEEE Transactions on Vehicular Technology, vol. 72, no. 8, pp. 10754-10769, Aug. 2023, doi: 10.1109/TVT.2023.3257107.

[17] P. Liu, K. An, J. Lei, Y. Sun, W. Liu and S. Chatzinotas, "Grant-Free SCMA Enhanced Mobile Edge Computing: Protocol Design and Performance Analysis," in IEEE Internet of Things Journal, 2024, doi: 10.1109/JIOT.2024.3386593.

[18] L. Wang, J. Xu, T. Qi, X. Jiang, J. Cui and B. Zheng, "An Optimization Method to Maximize the Service Quality of SCMA Grant-Free Access with MPR," 2021 13th International Conference on

Wireless Communications and Signal Processing (WCSP), Changsha, China, 2021, pp. 1-5, doi: 10.1109/WCSP52459.2021.9613275.

[19] M. J. Osborne, "An introduction to game theory," New York: Oxford university press, 2004.

[20] S. S. Abidrabbu and H. Arslan, "Energy-Efficient Resource Allocation for 5G Cognitive Radio NOMA Using Game Theory," 2021 IEEE Wireless Communications and Networking Conference (WCNC), 2021, pp. 1-5.

[21] M. Fadhil, A. H. Kelechi, R. Nordin, N. F. Abdullah and M. Ismail, "Game theory-based power allocation strategy for NOMA in 5G cooperative beamforming," in Wireless Personal Communications, 2022, 122(2): 1101-1128.

[22] R. Zheng, H. Wang, M. De Mari, M. Cui, X. Chu and T. Q. S. Quek, "Dynamic Computation Offloading in Ultra-Dense Networks Based on Mean Field Games," in IEEE Transactions on Wireless Communications, vol. 20, no. 10, pp. 6551-6565, Oct. 2021.

[23] J. M. Lasry and P. L. Lions, "Mean field games," Japanese Journal of Mathematics, vol. 2, no. 1, pp. 229–260, Mar. 2007.

[24] H. Gao et al., "Energy-Efficient Velocity Control for Massive Numbers of UAVs: A Mean Field Game Approach," in IEEE Transactions on Vehicular Technology, vol. 71, no. 6, pp. 6266-6278, June 2022.

[25] T. Li et al., "A mean field game-theoretic cross-layer optimization for multi-hop swarm UAV communications," in Journal of Communications and Networks, vol. 24, no. 1, pp. 68-82, Feb. 2022, doi:10.23919/JCN.2021.000035.

[26] M. De Mari, E. Calvanese Strinati, M. Debbah and T. Q. S. Quek, "Joint Stochastic Geometry and Mean Field Game Optimization for Energy-Efficient Proactive Scheduling in Ultra Dense Networks," in IEEE Transactions on Cognitive Communications and Networking, vol. 3, no. 4, pp. 766-781, Dec. 2017.

[27] A. Benamor, O. Habachi, I. Kammoun and J. -P. Cances, "Mean Field Game-Theoretic Framework for Distributed Power Control in Hybrid NOMA," in IEEE Transactions on Wireless Communications, vol. 21, no. 12, pp. 10502-10514, Dec. 2022.

[28] L. Li et al., "Resource Allocation for NOMA-MEC Systems in Ultra-Dense Networks: A Learning Aided Mean-Field Game Approach," in IEEE Transactions on Wireless Communications, vol. 20, no. 3, pp. 1487-1500, March 2021.

[29] R. S. Ganesan, W. Zirwas, B. Panzner, K. I. Pedersen and K. Valkealahti, "Integrating 3D Channel Model and Grid of Beams for 5G mMIMO System Level Simulations," 2016 IEEE 84th Vehicular Technology Conference (VTC-Fall), 2016, pp. 1-6.

[30] B. Wang, L. Dai, Z. Wang, N. Ge and S. Zhou, "Spectrum and Energy-Efficient Beamspace MIMO-NOMA for Millimeter-Wave Communications Using Lens Antenna Array," in IEEE Journal on Selected Areas in Communications, vol. 35, no. 10, pp. 2370-2382, Oct. 2017.

[31] 3GPP TS 36.211 V15.5.0, "Evolved universal terrestrial radio access (E-UTRA): Physical channels and modulation (Release 15)," Mar. 2019. [Online]. Available: https://portal.3gpp.org/desktopmodules/Specifications/SpecificationDetails.aspx?specificationId=2425. [Accessed: July 25, 2024]

[32] H. Chen, Y. Gu and S. -C. Liew, "Age-of-Information Dependent Random Access for Massive IoT Networks," IEEE INFOCOM 2020 - IEEE Conference on Computer Communications Workshops (INFOCOM WKSHPS), Toronto, ON, Canada, 2020, pp. 930-935.

[33] H. Khan, M. M. Butt, S. Samarakoon, P. Sehier and M. Bennis, "Deep Learning Assisted CSI Estimation for Joint URLLC and eMBB Resource Allocation," 2020 IEEE International Conference on Communications Workshops (ICC Workshops), Dublin, Ireland, 2020, pp. 1-6.

[34] M. M. Olama, S. M. Djouadi, and C. D. Charalambous, "Stochastic power control for time-varying long-term fading wireless networks," EURASIP Journal on Advances in Signal Processing, 2006.

[35] F. Tang, Y. Zhou and N. Kato, "Deep Reinforcement Learning for Dynamic Uplink/Downlink Resource Allocation in High Mobility 5G HetNet," in IEEE Journal on Selected Areas in Communications, vol. 38, no. 12, pp. 2773-2782, Dec. 2020

[36] S. Lasaulce and H. Tembine, "Game theory and learning for wireless networks : fundamentals and applications," Oxford ; Waltham, Ma: Academic Press, 2011. [Online]. Available: https://www.researchgate.net/publication/278768710_Game_Theory_and_Learning_for_Wireless_Networks_Fundamentals_and_Applications. [Accessed: July 25, 2024]

[37] R. Bellman, "Dynamic programming and stochastic control processes," Information and control, vol. 1, no. 3, pp. 228-239, 1958.

[38] Y. Jiang, Y. Hu, M. Bennis, F. Zheng and X. You, "A Mean Field Game-Based Distributed Edge Caching in Fog Radio Access Networks," in IEEE Transactions on Communications, vol. 68, no. 3, pp. 1567-1580, March 2020.

[39] T. Başar and G. J. Olsder, "Dynamic noncooperative game theory," Philadelphia, Pa: Society For Industrial And Applied Mathematics, 1999.

[40] X. Ge, H. Jia, Y. Zhong, Y. Xiao, Y. Li and B. Vucetic, "Energy Efficient Optimization of Wireless-Powered 5G Full Duplex Cellular Networks: A Mean Field Game Approach," in IEEE Transactions on Green Communications and Networking, vol. 3, no. 2, pp. 455-467, June 2019.

[41] B. Blaszczyszyn, M. Jovanovic, and M. K. Karray, "Performance laws of large heterogeneous cellular networks," in 2015 13th International Symposium on Modeling and Optimization in Mobile, Ad Hoc, and Wireless Networks (WiOpt), May 2015, pp. 597–604, 00001.

[42] M. De Mari and T. Quek, "Energy-efficient proactive scheduling in ultra dense networks," 2017 IEEE International Conference on Communications (ICC), Paris, 2017, pp. 1-6.

[43] M. Burger and J. M. Schulte, "Adjoint methods for HamiltonJacobi-Bellman equations," Munster, Germany:Univ. Munster, 2010.

**Appendix A**

As $v_n(t, x_n(t))$ is the value function of cost $C_n(t)$ at the state $X_n(t)$, according to Bellman's principle of optimality, increasing time $t$ to $t + dt$, leads to:

$$v_n(t, X_n(t)) = \min_{D_n(t)} \mathbb{E}\left[\int_t^{t+dt} C_n(u)du + v_n(t+dt, X_n(t+dt))\right] \quad (34)$$

By performing Taylor's expansion on $v_n(t, X_n(t))$, we get:

$$v_n(t+dt, X_n(t+dt)) = v_n(t, X_n(t)) + \partial_t v_n(t, X_n(t)) + \partial_t X_n(t) \cdot \nabla v_n(t, X_n(t))dt + o(dt) \quad (35)$$

Then, by substituting (35) into (34), subtracting $v_n(t, X_n(t))$ from both sides of the equation, and dividing both sides by $dt$. When $dt$ approaches zero, $o(dt)$ tends to zero and is negligible. Therefore (34) can be written as:

$$\min_{D_n(t)}\left[C_n(t) + \partial_t X_n(t) \cdot \nabla v_n(t, X_n(t))\right] + \partial_t v_n(t, X_n(t)) = 0 \quad (36)$$

Because of $X_n(t) = [E_n(t), \hat{H}_n(t)]$ and $C_n(t) = D_n^2(t)$, we obtain the HJB equation.

**Appendix B**

From (40), the optimal backoff delay $D_n^*(t)$ can be derived as:

$$\begin{aligned}D_n^*(t) = \arg\min_{D_n(t)}\Big[&-\frac{\gamma_0}{K \cdot l_n \cdot \hat{H}_{nm}^2(t)}[I_n(t, D_n(t)) + |w_{nm}^H(t)n_0|^2 B] \cdot \frac{\partial v^*(t, X_n(t))}{\partial E_n(t)} \\ &+ \delta_{nm}^a(t)|w_{nm}^H(t)\alpha(t, h_{nm}(t))|\frac{\partial v^*(t, X_n(t))}{\partial \hat{H}_{nm}(t)} \\ &+ \frac{\delta_{nm}^b(t)|w_{nm}^H(t)\sigma_{h,i}|^2}{2}\frac{\partial^2 v^*(t, X_n(t))}{\partial \hat{H}_{nm}^2(t)} + D_n^2(t)\Big]\end{aligned} \quad (37)$$

For the first derivative of the Hamiltonian with respect to $D_n(t)$:

$$\frac{\partial H_{am}}{\partial D_n(t)} = -\frac{\gamma_0}{K \cdot l_n \cdot \hat{H}_{nm}^2(t)}\frac{\partial I_n(t, D_n(t))}{\partial D_n(t)} \cdot \frac{\partial v^*(t, X_n(t))}{\partial E_n(t)} + 2D_n(t) \quad (38)$$

Taking the derivative of the interference term, we can obtain:

$$\frac{\partial I_n(t, D_n(t))}{\partial D_n(t)} = \frac{\beta}{|\Phi_m(t)| - 1}\mathbb{E}\{p_{n'}(t)\hat{H}_{n'm}(t)\} \cdot \frac{\partial |\Phi_m(t, D_n(t))|}{\partial D_n(t)} \quad (39)$$

In which, $|\Phi_m(t, D_n(t))|$ represents the number of devices whose backoff delay is $D_n(t)$ in cluster *m*. It has no explicit mathematical relationship with the backoff delay, but in the process of using the MFG to solve, $|\Phi_m(t, D_n(t))|$ can be converted from the mean field density, which is differentiable. So the partial differential equation of the interference term with respect to $D_n(t)$ exists. Therefore, Hamiltonian is smooth. The minimum value of $D_n(t)$ exists in which the first order partial derivative of Hamiltonian with respect to it is equal to zero, i.e.

$$-\frac{\gamma_0}{K \cdot l_n \cdot \hat{H}_{nm}^2(t)} \frac{\partial I_n(t, D_n(t))}{\partial D_n(t)} \cdot \frac{\partial v^*(t, X_n(t))}{\partial E_n(t)} + 2D_n(t) = 0 \tag{40}$$

Therefore, the backoff delay can be derived as (18).

**Appendix C**

The interference in (8) can be transformed into:

$$I_n(t, D_n(t)) = \mathbb{E}(l_{n'}) \sum_{n' \in \Phi_m(t, D_n(t)), n' \neq n} p_{n'}(t, D_n(t)) \hat{H}_{n'm}^2(t) \tag{41}$$

in which

$$\mathbb{E}(l_{n'}) = \frac{\int_0^{r_m} l_{n'} d(\pi r^2)}{\pi r_m^2} \tag{42}$$

in which $r_m$ is the radius of cluster m. Since the number of other devices in each cluster can be estimated by cell area and device density $\rho$ which satisfy $|\Phi_m(t)|-1 = \rho \cdot \pi r_m^2$. The interference can be derived as:

$$I_n(t) = \frac{\beta}{|\Phi_m(t)|-1} \sum_{n' \in \Phi_m(t, D_n(t)), n' \neq n} p_{n'}(t, D_n(t)) \hat{H}_{n'm}^2(t) \tag{43}$$

where $\beta = \rho\pi(1+\frac{2}{a-2}-r_m^{2-a})$.

**Appendix D**

Let's suppose a smooth and compactly supported function $y(X)$, It can be deduced that:

$$\int m(t, X) y(X) dX = \frac{1}{N} \sum_{n=1}^{N} y(X_n(t)) \tag{44}$$

By taking the partial derivative of t on both sides of the equation and applying the chain rule of derivation, we can get:

$$\int \partial_t m(t,X) y(X) dX \approx \frac{1}{N} \sum_{n=1}^{N} \left[ \partial_t X_n(t) \nabla y(X_n) + \partial_t^2 X_n(t) \Delta y(X_n) \right] \quad (45)$$

When n tends to infinity, (45) converts to:

$$\int \partial_t m(t,X) y(X) dX = \int \left[ \partial_t X_n(t) \nabla y(X_n) + \partial_t^2 X_n(t) \Delta y(X_n) \right] m(t,X) dX \quad (46)$$

Applying integration by parts on (46), convert it to:

$$\int \left[ \partial_t m(t,X) + \partial_t X(t) y(X) \nabla m(t,X) - \partial_t^2 X(t) \Delta m(t,X) \right] y(X) dX = 0 \quad (47)$$

When assuming $y(X)=1$, (47) can be converted to:

$$\partial_t m(t,X) + \partial_h (\alpha(t) m(t,X)) - \partial_E (p(t, D^*(t), X) m(t,X)) - \frac{1}{2} \sigma^2 \partial_{hh} m(t,X) = 0 \quad (48)$$

Since $p(t, D^*(t), X) = \frac{\gamma}{l \cdot \hat{H}^2(t)} \left( I(t, D^*(t)) + N_0 B \right)$, in which $P(t, D^*(t), X)$ is not affected by $E$ under the condition of given $D_n(t)$, according to the chain rule of derivation:

$$\frac{\partial p(t, D_n(t), X)}{\partial_E} = \frac{\partial p(t, D^*(t), X)}{\partial_D} \cdot \frac{\partial D^*(t)}{\partial_E} = \left( \frac{\gamma}{2K \cdot \hat{H}_m^2(t)} \frac{\partial I(t, D^*(t))}{\partial_D} \right)^2 \cdot \frac{\partial^2 v^*(t, X_n(t))}{\partial_{E^2}} \quad (49)$$

Since $\frac{\partial^2 v^*(t, X_n(t))}{\partial_{E^2}} = 0, \frac{\partial p(t, D_n(t), X)}{\partial_E} = 0$, the final form of the FPK equation can be derived as (32).